# Reply to Comment on "Space-Time Crystals of Trapped Ions"

Tongcang Li[1], Zhe-Xuan Gong[2,3], Zhang-Qi Yin[3,4], H. T. Quan[5], Xiaobo Yin[1], Peng Zhang[1],

L.-M. Duan[2,3], Xiang Zhang[1,6,*]

[1]NSF Nanoscale Science and Engineering Center, 3112 Etcheverry Hall, University of California, Berkeley, California 94720, USA
[2]Department of Physics, University of Michigan, Ann Arbor, Michigan 48109, USA
[3]Center for Quantum Information, Institute for Interdisciplinary Information Sciences, Tsinghua University, Beijing, 100084, P. R. China
[4]Key Laboratory of Quantum Information, University of Science and Technology of China, Chinese Academy of Sciences, Hefei, 230026, P. R. China
[5]Department of Chemistry and Biochemistry, University of Maryland, College Park, Maryland 20742, USA
[6]Materials Sciences Division, Lawrence Berkeley National Laboratory, 1 Cyclotron Road, Berkeley, California 94720, USA
*Corresponding author: xiang@berkeley.edu

The Comment [1] from Patrick Bruno (PB) shows a misunderstanding of our Letter [2]. Here we clarify several key concepts, and discuss the effects of pinning potentials on the persistent one-way rotation of trapped ions.

PB misinterpreted a quantum time crystal (QTC) [3] to be a system that has a periodic oscillation of some physical observable in its *absolute* quantum ground state. Actually, we clearly stated at the 8th line of the first paragraph on page 1 and the whole paragraph from the bottom of the right column of page 3 to the top of page 4 in our letter [2] that the rotating inhomogeneous state of a QTC is an *effective* ground state. For example, starting at the 9th line on page 4, we said: "The resulted rotating inhomogeneous state is *not the true ground state, but an effective ground state* with its energy infinitely close to that of the true ground state when $N \to \infty$". This is analogous to the fact that a spatial crystal in nature formed after translational symmetry breaking is in an effective ground state [4, 5]. For a Hamiltonian with spatial translational invariance, as pointed out by Anderson at the 4th line on page 39 of Ref. [5], the eigenstates "*are not position eigenstates but momentum eigenstates, devoid of locality* *information*". However, a spatial crystal localizes at specific spatial positions [4, 5]. So it cannot be in the true ground state of the spatial invariant Hamiltonian, but in an effective ground state that is a combination of eigenstates [4, 5]. From the 13th line on page 4 of our letter [2], we also *gave a practical way* to create the desired effective ground state in our *finite* experimental system: "a weak observation of a single particle can localize the center of mass of all particles and project the true ground state to a rotating effective ground state".

Another question PB raised is the radiation loss. The *true* ground state of the Hamiltonian, as discussed in the last paragraph of page 2 of our letter [2], has no radiation loss. The *effective* ground state after the symmetry breaking will have radiation loss, but it is negligible. The Larmor formula of electrodynamics predicts the radiation-limited lifetime of the rotation of a $^9Be^+$ ion in a 100-micron-diameter ring at 0.4 Hz studied in [2] to be on the order of $10^{18}$ years, which is irrelevant to an experiment. In fact, the effective ground states of spatial crystals are also not absolutely stable. As stated at the 1st line of the 3rd page of Ref. [4], which was also cited by Bruno [1], "the *delocalization* of the crystal due to the spreading of its wavefunction will take a time proportional to $N$ and can thus never be observed".

Furthermore, we point out here that PB used an assumption different from ours when he examined the effects of the order of taking the limits of the particle number $N \to \infty$ and the pinning potential $v \to 0$. PB assumed the lattice spacing to be constant $a$, and found $\lim_{v \to 0} \lim_{N \to \infty} A_a(v,N) = 0$, where $A_a(v,N) \approx \exp(-\frac{a}{\hbar}\sqrt{2NMv})$ is a tunneling factor of an ion crystal through an effective pinning potential $v$ [6], and $M$ is the mass of each ion. Under PB's assumption, it is not surprising that the ion crystal will not rotate when $N \to \infty$ because the size of the ion ring will be infinite. However, as described in our work [2], we considered a more interesting situation by assuming the diameter of the ion ring to be a constant $d$, which is similar to the case of electrons in a metal (or semiconductor) ring with a fixed size. Then $a = \pi d / N$ and $A_d(v,N) \approx \exp(-\frac{\pi d}{\hbar}\sqrt{2Mv/N})$. It is easy to see that $\lim_{v \to 0} \lim_{N \to \infty} A_d(v,N) = 1$ and $\lim_{N \to \infty} \lim_{v \to 0} A_d(v,N) = 1$, independent of the order of taking the limit. PB's infinite $N$ limit at a fixed-density and our infinite $N$ limit at a fixed-size correspond to different situations. For a spatial crystal found in nature, the distance between the atoms is almost a constant. The size of a spatial crystal increases when it grows. So it is natural to take the fixed-density limit for natural spatial crystals. For ions in a ring trap, however, the size is determined by the apparatus of the ion trap and is almost a constant. The separation between ions will change when we load more ions to the trap. So it is natural to choose the fixed-size limit for trapped ions. Our proposal provides an experimental scheme to observe the formation process of time crystal as ion number increases (yet still finite). Strictly speaking, PB's fixed-density limit would require a spatial crystal to have an infinite size, while our fixed-size limit would require a time crystal to have an infinite density. In reality, all spatial crystals found in nature have finite sizes, and the spatial symmetry breaking still occurs naturally because the pinning potential from the environment is small but finite. Similarly, although it is unphysical to trap an infinite number of ions with a fixed-size trap, the temporal symmetry breaking can occur naturally and a finite time crystal will be formed with a small time-dependent perturbation. For a moderate number of ions, e.g. N=100, if the environmental perturbations are from the scattering of photons as described in [2] and each photon scattering causes 5% momentum uncertainty in the ring direction, then the required average number of photons scattered by each ion will be on the order of 0.01, which is already rather small. As a result, the emergence of time crystal is a physical consequence in the proposed experiment, even though it has a finite size.